\documentclass[aps,prl]{revtex4}%
\usepackage{amsfonts}
\usepackage{amsmath}
\usepackage{amssymb}
\usepackage{graphicx}%
\providecommand{\U}[1]{\protect\rule{.1in}{.1in}}

\providecommand{\U}[1]{\protect\rule{.1in}{.1in}}
\providecommand{\U}[1]{\protect\rule{.1in}{.1in}}

\begin{document}
\preprint{ }
\title[Short title for running header]{Enhancement of ultrafast electron photoemission from metallic nanoantennas
excited by a femtosecond laser pulse }
\author{M. A. Gubko$^{1}$, W. Husinsky$^{2}$, A. A. Ionin$^{1}$, S. I.
Kudryashov$^{1\ast}$, S. V. Makarov$^{1,2\ast}$,C.S.R. Nathala$^{2}$, A. A.
Rudenko$^{1,}$ L. V. Seleznev$^{1}$, D. V. Sinitsyn$^{1}$, I.V. Treshin$^{1}$ }
\affiliation{$^{1}$P.N.Lebedev Physical Institute of the Russian Academy of Sciences,
119991 Leninsky pr. 53, Moscow}
\affiliation{$^{2}$Vienna University of Technology, 101040 Wiedner Hauptstrasse 8, Vienna, Austria}

\begin{abstract}
We have demonstrated for the first time that an array of nanoantennas (central
nanotips inside sub-micron pits) on an aluminum surface, fabricated using a
specific double-pulse femtosecond laser irradiation scheme, results in a
28-fold enhancement of the non-linear (three-photon) electron photoemission
yield, driven by a third intense IR femtosecond laser pulse. The supporting
numerical electrodynamic modeling indicates that the electron emission is
increased not owing to a larger effective aluminum surface, but due to instant
local electromagnetic field enhancement near the nanoantenna, contributed by
both the tip's \textquotedblleft lightning rod\textquotedblright\ effect and
the focusing effect of the pit as a microreflector and annular edge as a
plasmonic lens.

\end{abstract}
\volumeyear{year}
\volumenumber{number}
\issuenumber{number}
\eid{identifier}
\date[Date text]{date}
\received[Received text]{date}

\revised[Revised text]{date}

\accepted[Accepted text]{date}

\published[Published text]{date}

\startpage{1}
\endpage{ }
\maketitle

Strong-field plasmonics, involving excitation of plasmons (collective
free-electrons oscillations) in different nanoobjects by intense femtosecond
(fs) laser pulses, is of high interest for basic and applied research.
Surface-plasmon-enhanced multi-photon photoelectric emission \cite{01},
high-harmonic generation \cite{02}, electron acceleration \cite{03,04} and
x-ray enhancement \cite{05} were demonstrated using such nanostructures as
diffractive gratings \cite{03}, plasmonic bow-tie nanoantennas\cite{02,04},
spherical \cite{06} and nonspherical \cite{04,05} metallic nanoparticles.

One of the most popular plasmonic elements is a metallic nanotip, providing
strong optical field enhancement via the \textquotedblleft lightning rod
effect\textquotedblright. The nanometer-long decay length of the evanescent
field corresponds to its strong gradients, which can be used for nanoscale
acceleration of photo-emitted electrons in different regimes (multiphoton
\cite{07}, above-threshold \cite{08} or optical-field \cite{09} regimes).
Interestingly, that the strong gradient of localized evanescent field can
suppress the quiver motion of the electrons in the oscillating laser electric
field \cite{10}. Such a strong-field steering of electrons in the vicinity of
nanostructures with large local field enhancement and steep field gradients
leads to emission of highly-directed, confined coherent electron wavepackets
\cite{07,09,10,11}. Generally, such a pulsed electron nanoemitter, triggered
by femtosecond laser irradiation, could serve as an efficient source for
time-resolved nanoscale imaging. For instance, ultrashort electron pulses were
employed for time-domain visualization of metal melting \cite{12} and
ionization dynamics of H${}_{2}$ \cite{13}.

Fabrication of plasmonic nanotips usually faces problems of long fabrication
cycle, chemical treatment and production costs. To provide more efficient
fabrication ways, tight focusing of single nanosecond \cite{14} and
femtosecond \cite{15} laser pulses into diffraction-limited spots was tested
to produce one nanotip per shot. However, femtosecond laser irradiation makes
it possible and realistic to easily fabricate huge arrays of nanostructures
(down to the sub-100-nm scale) via intense surface plasmons polaritons (SPPs)
excitation, where only weak laser beam focusing on the surface is required
\cite{16}. Such a method for surface nanograting formation was i.e.
successfully used for surface-plasmon-enhanced photoelectron emission
\cite{01}. In the same manner, also an array of nanotips can be easily
fabricated by means of fs-laser beam weak focusing on a metallic surface
\cite{17}.

In this Letter, we report a simple, double-pulse fs-laser fabrication scheme
to produce an array of nanoantennas (nanotips inside sub-micron pits) on an
aluminum surface and demonstrate their strongly enhanced non-linear electron
photoemission, excited by a single fs-laser pulse, in comparison to flat and
randomly nanostructured aluminum surfaces. These observations are supported by
numerical electrodynamic modeling, indicating high local electromagnetic (EM)
field enhancement in the nanoantennas.

In our experiments 100-fs, 744-nm linearly-polarized Ti:sapphire laser pulses
with a maximum pulse energy of 6 mJ in the TEM${}_{00}$-mode were focused by a
silica lens (focal distance of 11 cm) onto a 4-mm-thick aluminum sample
mounted vertically on an \textit{X}-\textit{Y-Z }motorized translation stage.
The mechanically polished and ultrasonically cleaned sample was located
several mm above the focal plane to obtain a large spot diameter
\textit{D${}_{1/e}$}${}\thickapprox$ 180 mm. The nanostructured samples
surfaces were characterized using field-emission scanning electron microscopy (FE-SEM).

To measure photoelectron emission, a stationary collecting aluminum electrode
(anode) with a 2-mm aperture was mounted at a distance of 1 mm away from the
sample surface and a positive voltage of 150 V was applied to extract the
emitted electrons (the scheme was described elsewhere \cite{18}). The fs-laser
pump pulses were focused on the target surface through the anode aperture. The
extracting field ($\thickapprox$1 kV/cm) in this scheme is two or three orders
of magnitude higher than values typical for high-vacuum schemes, where the
field values must not exceed $\thickapprox$1--10 V/cm to prevent secondary
electron emission, since at atmospheric pressure emitted electrons become
attached to oxygen molecules on a nanosecond time scale. Then, the resulting
negatively charged ions slowly move in the applied electric field on a
sub-millisecond time scale, inducing an image current (potential) in the
collector, which was registered using a M$\Omega$-input of a digital
oscilloscope. The high extracting electric fields eliminate the space-charge
effect even at intense electron emission at fs-laser fluences even as high as
several J/cm${}^{2}$ \cite{18}.

Nanoantennas fabrication on an aluminum surface was performed by two fs-laser
pulses at the same peak fluence \textit{F${}_{0}$} $\thickapprox$ 0.85
J/cm${}^{2}$ (slightly below the spallative ablation threshold \textit{F${}%
_{spal}$} $\thickapprox$ 0.7 J/cm${}^{2}$ \cite{19}), following with a delay
of a few seconds between them \cite{17}. After the first laser pulse an
irregular array of round spallative pits with a surface density $\sim$%
10${}^{7}$ cm${}^{-2}$ appeared on the surface (Fig. 1a) at local fluences
\textit{F} $>$ \textit{F${}_{spal}$} along an outer border of a macroscopic
spallation crater. Their edges have widths of about $\Delta\thickapprox$ 100
nm, their bottom is semispherical appearing, in average, 100 nm below the
initial surface level (Fig. 1b). The average diameter of the pits depends on
local laser fluence, but usually amounts to 1.3 $\mu$m. They result from
intense sub-surface nanovoid generation (homogeneous nucleation) in the melted
surface layer \cite{20,21} at fs-laser fluences slightly lower than the
spallation threshold \textit{F${}_{s}$}.

%

\begin{figure}
[ptbh]
\begin{center}
\includegraphics[
width=2.3367in
]%
{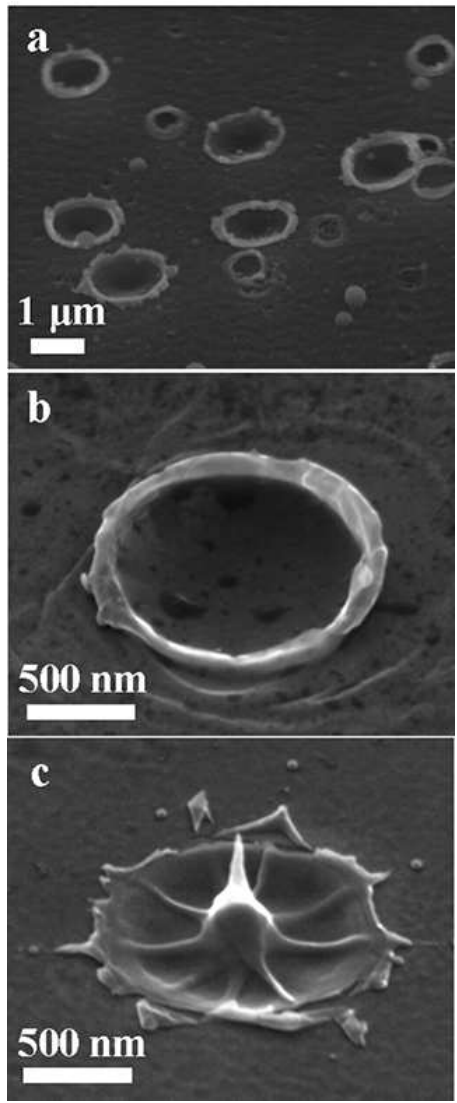}%
\caption{ FE-SEM images of a pit array (a) and a separate pit (b) on aluminum
surface produced by single fs-laser pulse at\textit{\ F${}_{0}$}
$\thickapprox$ 0.85 J/cm${}^{2}$. (c) FE-SEM image of the pit upon its
exposure by the second fs-laser pulse at this fluence.}%
\end{center}
\end{figure}

Such pits with prominent edges respond to EM fields in the optical range as
plasmonic nanolenses \cite{22}, providing excitation and sub-diffraction
focusing of SPPs in their centers. The focusing in plasmonic lenses exposed by
fs-laser pulses at \textit{F${}_{0}$} $\thickapprox$ 0.85 J/cm${}^{2}$ results
within each pit in the formation of a single nanojet (Fig. 1c), related to
material expulsion and its ultrafast cooling \cite{17} expected for much
higher fs-laser fluences, exceeding the threshold \textit{F${}_{frag}$}
$\thickapprox$1.4 J/cm${}^{2}$ for supercritical hydrodynamic (fragmentation)
\cite{19}.

\noindent

To evaluate the optical field enhancement in such a nanoantenna (a nanojet in
a microscale pit), we performed numerical modeling by solving Maxwell's
equations using finite-elements method (COMSOL). EM intensity distribution was
calculated for a plane EM wave ($\lambda$ = 744 nm reaching a nanotip in a pit
at normal incidence (Fig. 2a) with the geometrical parameters: \textit{H} =
550 nm, \textit{h }= 100 nm, \textit{R${}_{0}$} = 650 nm, \textit{R} = 100 nm,
\textit{r} = 20 nm, $\Delta$ = 100 nm (notations see in Fig. 2a), taken from
Fig. 1c. The dielectric function of unexcited aluminum at the 744-nm
wavelength equals $\mathit{\varepsilon=}-68.9\mathit{+i}39.9$ \cite{23}.

%

\begin{figure}
[ptbh]
\begin{center}
\includegraphics[
width=2.9162in
]%
{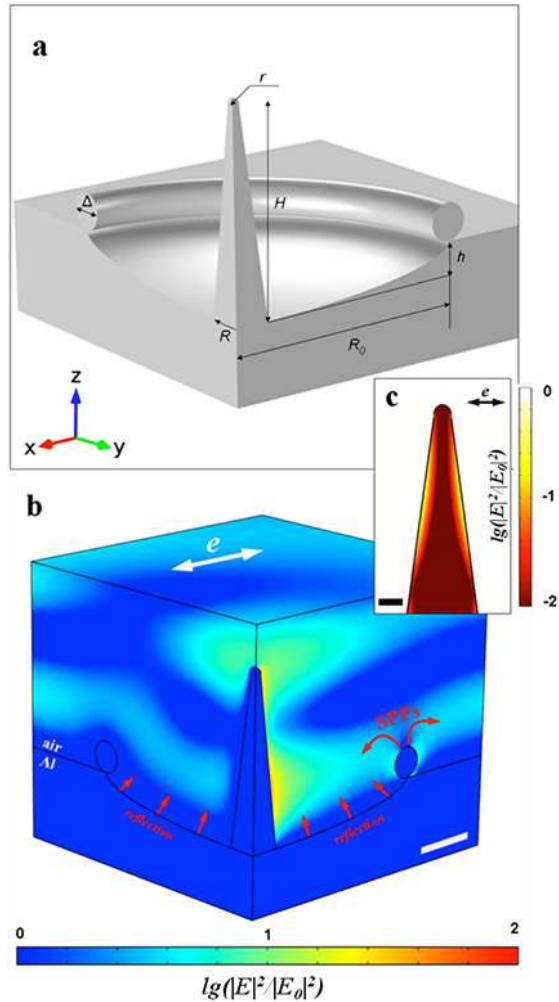}%
\caption{(a) Cross-section view of a 3D-model nanotip in a pit on aluminum
surface with the shown notations of their geometrical parameters. (b)
Calculated decimal logarithm of squared field enhancement \textit{lg}%
($\vert$\textit{E}$\vert$${}^{2}$/$\vert$\textit{E${}_{0}$}$\vert$${}^{2}$)
distribution near the model structure with \textit{H} = 550 nm, \textit{h} =
100 nm, \textit{R }= 100 nm, \textit{R${}_{0}$} = 650 nm, \textit{r} = 20 nm,
$\Delta$ = 100 nm. (c) Image of the nanotip with the internal intensity
distribution. Double arrows in pictures (b) and (c) indicate orientation of
the EM wave linear polarization direction. White scale in (b) indicates 200 nm
length. Black scale in (c) indicates 30 nm.}%
\end{center}
\end{figure}

This modeling has revealed an intensity enhancement up to 56 times outside and
5.5 times inside the peak of the nanotip (Fig. 2b,c). The enhancement factor
inside the nanotip is the ratio between the maximal laser intensity values
under the nanotip surface and under the flat metallic surface. The model
calculation takes into account all possible interference effects, and,
consequently, the enhancement is attributed not only to local phenomena such
as the ``lighting rod'' effect, but also to EM wave reflection from the
semi-spherical surface of the pit and SPP excitation from its edges.
Calculation of the field for a nanotip on a flat aluminum surface resulted in
a corresponding local field enhancement factor 2 times lower (as compared to a
nanotip in a pit) outside the nanotip and 1.3 times lower inside. Hence, this
proves that the pit works like a reflector in a parabolic antenna, which
focuses the incident EM waves onto the nanotip. Additionally, in our case such
pits have sharp edges, providing SPPs excitation and focusing to the nanotip.

\noindent

To study the electron emissivity of the fabricated array of the
fs-laser-induced nano-tips in the micro-craters, we measured the photoemission
of electrons from the nano-structured surface in the appropriate intensity
regime ($\thickapprox$1-10 TW/cm${}^{2}$), where the micro-craters and
nano-tips are typically formed, and compared the yield from a polished surface
with the yield from surfaces with laser-induced random nanostructures.%

\begin{figure}
[ptbh]
\begin{center}
\includegraphics[
height=5.7908in,
]%
{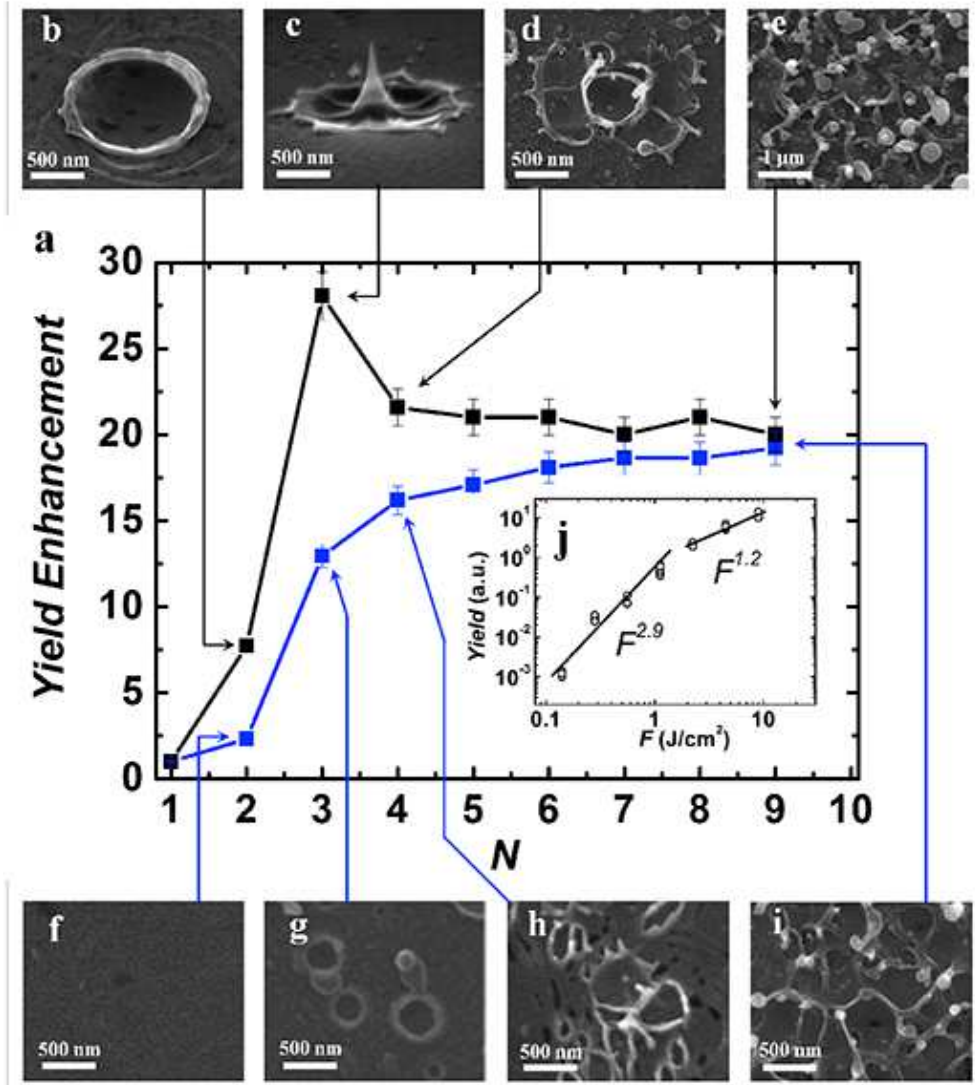}%
\caption{ (a) Evolution of the electron yield enhancement with the number of
fs-laser shots at \textit{F${}_{0}$} = 0.5 J/cm${}^{2}$ (blue curve) and 0.85
J/cm${}^{2}$ (black curve). The characteristic surface nanofeatures
\textit{before} irradiation by the second (b,f), third (c,g), fourth (d,h) and
ninth (e,i) fs-laser pulse at \textit{F${}_{0}$} = 0.5 J/cm${}^{2}$ (b-e) and
0.85 J/cm${}^{2}$ (f-i). (j) Single-shot electron emission yield dependence on
fluence for the reference flat aluminum surface.}%
\end{center}
\end{figure}

In Fig. 3 the enhancement of the electron photoemission is shown versus
\textit{N} at two fluences \textit{F${}_{0}$} $\thickapprox$ 0.85 J/cm${}^{2}$
and \textit{F${}_{0}$} $\thickapprox$ 0.5 J/cm${}^{2}$. At the highest fluence
\textit{F${}_{0}$} $\thickapprox$ 0.85 J/cm${}^{2}$, the electron yield
enhancement is characterized by a maximum of $\thickapprox$ 28 at \textit{N} =
3 and subsequent decrease for increasing laser exposure \textit{N} $>$ 3,
since the nanotips are destroyed in the next shot, leaving nanopits underneath
them within the sub-micron pits (Fig. 3d). The succeeding multi-shot fs-laser
exposure results in a random structure of ablative nanoparticles (Fig. 3e),
providing the saturated electron yield enhancement about 20.

Similar multi-shot electron yield enhancement is achieved for smaller fluence
\textit{F${}_{0}$} $\thickapprox$ 0.5 J/cm${}^{2}$, which does not produce
high-fluence nanotips, but just lower-fluence nanopits (Fig. 3g). Eventually,
multi-shot irradiation in this fluence regime leads in similar random
nanorelief (Fig. 3i) through cumulative random formation of multiple
overlapping surface nanopits via the sub-surface cavitation mechanism [20,21].
As a result, for large \textit{N} the surface is again covered by
nanoparticles (Fig. 3i) due to enhanced local ablation in the nanopits (Fig.
3h). In this case, the electron yield enhancement factor grows monotonically
up to the almost same saturation level of $\thickapprox$ 20.

\noindent

The fs-laser induced electron emission enhancement factor of nearly 30
achieved for the nanoantennas has a straightforward explanation in terms of
the local field enhancement in the nanofeature. For that purpose, we have
obtained the experimental dependence of the electron emission yield on fluence
for the flat Al surface. The variation of electron emission yield as a
function of \textit{F${}_{0}$} is represented by the consequent cubic and
linear dependences for \textit{F${}_{0}$} $<$ 1.5 J/cm${}^{2}$ and
\textit{F${}_{0}$} $>$ 1.5 J/cm${}^{2}$, respectively (Fig. 3j). This
indicates that, for the incident fs-laser fluence \textit{F${}_{0}$}
$\thickapprox$ 0.85 J/cm${}^{2}$ the electron emission yield from the
reference flat surface exhibits in Fig. 3j magnitudes of 0.2-0.4 arbitrary
units within the third-power region of its fluence dependence. Following the
local intensity enhancement of 5.5 inside the nanotip, the effective fluence
becomes equal to 4-5 J/cm${}^{2}$, corresponding to the electron emission
yield values of 5-8 arbitrary units within the linear region of its fluence
dependence (Fig. 3j). As a result, we would expect an enhancement of the
photoemission yield due to the nanotips in range 15 -- 40. However, the
surface after the second fs-pulse is covered by nanotips only in part (less
than 10\% of the irradiated surface). In this case total electron
photoemission yield has contribution from excitation of SPPs outside the pits,
where they interfere with the incident laser field and each other on
relatively large area. Such interference SPP-light is the main origin of the
yield enhancement in case of \textit{N} = 3 at \textit{F${}_{0}$}
$\thickapprox$ 0.5 J/cm${}^{2}$ (Fig. 3g), where the rare sub-wavelength pits
play role of SPPs sources. It should be noted, that in comparison with random
nanostructures, a surface with nanotips has evidently a smaller density of
nanoelements, but a higher electron emission yield, indicating even stronger
local EM field enhancement on individual nanotips.

Moreover, another important characteristic of the nanoantennas is their large
($\thickapprox$50) electrodynamical enhancement of optical intensity outside
the nanotip, which is significantly higher than the internal enhancement
factor of $\thickapprox$5 inside (Fig. 2). Such discrepancy between both
enhancements results from their different electrical field polarizations.
Particularly, the internal electric field inside the nanotip appears as a
mostly longitudinal one with the predominating \textit{E${}_{z}$}-component
(Fig. 2c), as compared to the external electric field near the nanotip apex
with nearly equal \textit{E${}_{x}$}- and \textit{E${}_{z}$}-components (Fig.
2b). The EM wave reflected for the pit bottom at the almost normal incidence
angle contributes presumably its transversal component (\textit{E${}_{x}$$>$}
\textit{E${}_{z}$}) to the nanotip apex field. Hence, the internal field
inside the nanotip apex is contributed by SPP waves with the predominating
\textit{E${}_{z}$}-component, which are rather inefficiently excited at the
pit edges.

As a result of such a high external EM field enhancement, such nanoantenna
design, accompanied by the related \textquotedblleft chemical
enhancement\textquotedblright\ effect of electronic structure of noble metals,
can be very promising for diverse nanophotonic applications, such as
surface-enhanced absorption \cite{24}, Raman scattering \cite{25} and
luminescence \cite{26}.

In conclusion, we have demonstrated for the first time that an array of
laser-induced metallic nanotips within semispherical sub-micron pits provides
28-fold enhancement of ultrafast electron photoemission. Numerical
calculations of the intensity distribution near a nanotip have proven that
such an assembly works like nanoantennas with microreflectors, yielding in
high EM field concentration near the peak of the nanotip. Comparative study of
electron emission from the nanotips versus other types of laser-induced
nanotopologies showed that the nanotips provide the highest enhancement,
despite relatively low surface density. The experiments were carried out at
intensities higher than the damage threshold for the nanotips to show that
their simple way of fabrication opens a possibility of their using in
high-fluence ($>$ 1 J/cm${}^{2}$) regime.

\noindent This work was partly supported by \textquotedblleft Die
\"{O}sterreichische Forschungsf\"{o}rderungsgesellschaft FFG \textquotedblleft
project SLFNM 834325\textquotedblright, Russian Foundation for Basic Research
(projects nos. 11-02-01202-a, 11-08-01165-a, 12-02-13506 ofi\_m\_RA,
12-20-33045 mol-a\_ved, 13-02-00971-a, 14-02-00460-a, 14-02-00748-a, and
14-02-00881-a), and by RAS Presidium's programs (nos. 13 and 24).

\bigskip

\bigskip

\bigskip

\noindent

\begin{thebibliography}{99}                                                                                               %


\bibitem {01}\noindent T. Y. Hwang, A. Y. Vorobyev, and C. Guo. Phys. Rev. B
\textbf{79}, 085425 (2009).\label{01}

\bibitem {02}S. Kim \textit{et al}. Nature \textbf{453}, 757 (2008).\label{02}

\bibitem {03}\noindent J. Kupersztych, P. Monchicourt, and M. Raynaud. Phys.
Rev. Lett. \textbf{86}, 5180 (2001).\label{03}

\bibitem {04}\noindent P. Dombi \textit{et al}. Nano Lett. \textbf{13}, 674
(2013).\label{04}

\bibitem {05}P. P. Rajeev \textit{et al.} Phys. Rev. Lett. \textbf{90}, 115002
(2003).\label{05}

\bibitem {06}\noindent A. Gloskovskii \textit{et al}. Phys. Rev. B
\textbf{77},\textbf{\ }195427\textbf{\ }(2008).\label{06}

\bibitem {07}\noindent C. Ropers \textit{et al}. Phys. Rev. Lett. \textbf{98},
043907 (2007).\label{07}

\bibitem {08}\noindent M. Schenk, M. Kr\"{u}ger, and P. Hommelhoff. Phys. Rev.
Lett. \textbf{105}, 257601 (2010).\label{08}

\bibitem {09}\noindent P. Hommelhoff \textit{et al. }Phys. Rev. Lett.
\textbf{96}, 077401 (2006).\label{09}

\bibitem {10}\noindent\ G. Herink \textit{et al}. Nature \textbf{483}, 190
(2012).\label{10}

\bibitem {11}\noindent D. J. Park \textit{et al.} Phys. Rev. Lett.
\textbf{109}, 244803 (2012).\label{11}

\bibitem {12}\noindent B. J. Siwick\textit{\ et al. }Science \textbf{302},
1382 (2003).\label{12}

\bibitem {13}H. Niikura \textit{et al.} Nature \textbf{417}, 917
(2002).\label{13}

\bibitem {14}\noindent J. P. Moening \textit{et al.} Appl. Phys. A
\textbf{95}, 635 (2009).\label{14}

\bibitem {15}F. Korte, J. Koch, and B. N. Chichkov. Appl. Phys. A \textbf{79},
879 (2004).\label{15}

\bibitem {16}\noindent E. V. Golosov \textit{et al.} Phys. Rev. B \textbf{83},
115426 (2011).\label{16}

\bibitem {17}\noindent M. A. Gubko \textit{et al.} JETP Lett. \textbf{97}, 599
(2013).\label{17}

\bibitem {18}\noindent A. A. Ionin \textit{et al.} JETP Lett. \textbf{96}, 375
(2012).\label{18}

\bibitem {19}\noindent A. A. Ionin \textit{et al.} JETP Lett. \textbf{94}, 34
(2011).\label{19}

\bibitem {20}E. Leveugle, D. S. Ivanov, and L. V. Zhigilei. Appl. Phys. A
\textbf{79}, 1643 (2004).\label{20}

\bibitem {21}A. A. Ionin \textit{et al.} JETP \textbf{116}, 347
(2013).\label{21}

\bibitem {22}Z. Liu \textit{et al.} Nano Lett. \textbf{5}, 91726
(2005).\label{22}

\bibitem {23}E. D. Palik, \textit{Handbook of Optical Constants of Solids}
(Academic, London, 1985).\label{23}

\bibitem {24}D. M. Schaadt, B. Feng, and E. T. Yu. Appl. Phys. Lett.
\textbf{86}, 063106 (2005).\label{24}

\bibitem {25}M. Moskovits. Rev. Mod. Phys. \textbf{57}, 783 (1985).\label{25}

\bibitem {26}T. Liebermann and W. Knoll. Coll. and Surf. A \textbf{171}, 115
(2000).\label{26}
\end{thebibliography}
\end{document}